\documentclass[12pt]{article}

\usepackage{epsfig}

\def\And{{\rm and\ }}
\newif\ifboo \boofalse
\def\Review#1{\boofalse{\it #1},}
\def\Name#1{{#1},}
\def\Vol#1{\ifboo Vol. {\bf #1}\else{\bf #1}\fi}
\def\Year#1{\ifboo #1\else(#1)\fi}
\def\Book#1{\bootrue{\it #1},}
\def\Page#1{\ifboo {\rm p. #1}\else{\rm #1}\fi}

\newcommand{\centps}[2]{
	\begin{center}
		\epsfig{file=#1,width=#2mm}
	\end{center}
}

\begin{document}

\setcounter{page}{1}
\pagestyle{plain}
\setcounter{equation}{0}

\begin{center}
	\begin{large}
		{\bf Spontaneous Jamming in One-Dimensional Systems\\}
	\end{large}
	\vspace{5mm}
	%\begin{large}
		O.J.~O'Loan, M.R.~Evans and M.E.~Cates\\
	%\end{large}
	\vspace{3mm}
	{\it Department of Physics and Astronomy, University of Edinburgh,\\
	Mayfield Road, Edinburgh EH9 3JZ, U.K.\\}
\end{center}

\vspace{5mm}

%###############################################################################
% Abstract
%###############################################################################

\begin{abstract}
We study the phenomenon of jamming in driven diffusive systems. We introduce a
simple microscopic model in which jamming of a conserved driven species is
mediated by the presence of a non-conserved quantity, causing an {\it
effective} long range interaction of the driven species. We study the model
analytically and numerically, providing strong evidence that jamming occurs;
however, this proceeds via a strict phase transition (with spontaneous symmetry
breaking) only in a prescribed limit. Outside this limit, the nearby transition
(characterised by an essential singularity) induces sharp crossovers and
transient coarsening phenomena. We discuss the relevance of the model to two
physical situations: the clustering of buses, and the clogging of a suspension
forced along a pipe.
\end{abstract}

\vspace{3mm}

\noindent
PACS numbers: 05.70Ln; 64.60-i; 89.40+k 

\vspace{7mm}

Many non-equilibrium physical situations can be modelled as driven
diffusive systems\cite{SZ}. An intriguing feature of certain driven systems is
their propensity to jam -- in traffic flow\cite{Nagel} jamming behaviour is a
fact of modern life and in colloid rheology the phenomenon of shear thickening
(dilatancy) is widely studied\cite{FMB}.

One-dimensional ($1d$) driven systems exhibit a wide variety of interesting
phenomena, including phase transitions and spontaneous symmetry
breaking\cite{EFGM}, which are precluded from $1d$ equilibrium systems (in the
absence of long range interactions). This suggests that the physics of jamming
might be captured in simple $1d$ models. In previous studies of simple $1d$
non-equilibrium models, jamming arises because of the presence of disorder or
inhomogeneities such as defect sites \cite{defect}. In contrast, the model we
introduce below is homogeneous; the jamming emerges via spontaneous symmetry
breaking. Jamming arises through a mechanism in which a non-conserved quantity
in the dynamics mediates an {\it effective} long range interaction of a
conserved quantity (driven species), even though the microscopic dynamics is
{\it local} and stochastic.

We now define the microscopic model we study, which we refer to as the Bus
Route Model (BRM) for reasons to become clear. The BRM is defined on a $1d$
periodic lattice with $L$ sites. Site $i$ has two variables $\tau_i$ and
$\phi_i$ associated with it, each of which can be either $1$ or $0$. When a
site is occupied by a ``bus'', $\tau_i$ is $1$ and if $\phi_i$ is $1$ the site
is said to have ``passengers'' on it\cite{MultiPassengers}; $\tau_i$ and
$\phi_i$ cannot both be $1$ simultaneously. There are $M$ buses in total and
the bus density $\rho = M/L$ is a conserved quantity. However, the total number
of sites with passengers is {\em not} conserved.

In updating the system, a site $i$ is chosen at random. If both $\tau_i$ and
$\phi_i$ are $0$, then $\phi_i \to 1$ with probability $\lambda$. If $\tau_i =
1$ and $\tau_{i+1} = 0$, then the bus at site $i$ hops forward with probability
$1 - (1-\beta)\phi_{i+1}$. If the bus hops, $\phi_{i+1}$ becomes $0$. Thus, a
bus hops with probability $1$ onto a site without passengers, and probability
$\beta$ onto a site with passengers thereby removing them. The probability that
passengers arrive at an empty site is $\lambda$.  We generally take $\beta <
1$, reflecting the fact that buses are slowed down by having to pick up
passengers. Buses are forbidden from overtaking each other but relaxing this
condition will have no significant effect\cite{Overtaking}. We remark that the
dynamics is local and does not satisfy detailed balance.

At this point it is useful to discuss two scenarios which illustrate possible
applications of the model and highlight the r\^oles of the conserved and
non-conserved quantities. The first and most obvious example is that of buses
moving along a bus-route.  Clearly, the ideal situation is that the buses are
evenly spaced so that they pick up roughly equal numbers of passengers.
However, what commonly occurs is that a bus falls behind the one in front and
consequently has more passengers awaiting it. Thus the bus becomes further
delayed and at the same time, following buses catch up with it, leading to a
cluster of buses. The number of passengers awaiting a bus gives an indication
of the elapsed time since the last bus went past and in this way communicates
information between the two buses, resulting in an effective long range
interaction\cite{RealBuses}.

We now turn to an alternative interpretation of the model describing a system
of driven particles, each of which can exist in two states of mobility. Each
time a bus hops to the right in the BRM, a vacancy moves to the left.  In the
new interpretation of the model, which can be thought of as the dual of the
BRM, the vacancies become ``particles'' and the non-conserved variable is the
mobility (hopping probability) of a particle, which is either $1$ or $\beta$
\cite{ExactMapping}. A possible application of this dual model is to the 
phenomenon of clogging. A simple scenario is the flow of particles suspended in
a fluid being forced through a pipe. The pipe is narrow enough to prevent the
particles passing each other and stationary particles may become weakly
attached to the pipe (with rate $\lambda$), reducing their mobility from $1$ to
$\beta$. At high density, individual particles move more slowly and therefore
are more likely to become attached to the pipe, thus impeding the motion of the
following particles and encouraging them to attach. Hence clogging ensues.
Although set up as a strictly $1d$ model (requiring the diameters of the
particles and the pipe to be comparable), a similar scenario could affect the
flow of any heterogeneous material with a tendency to solidify when at
rest\cite{Galilean}.

\begin{figure}
	\centps{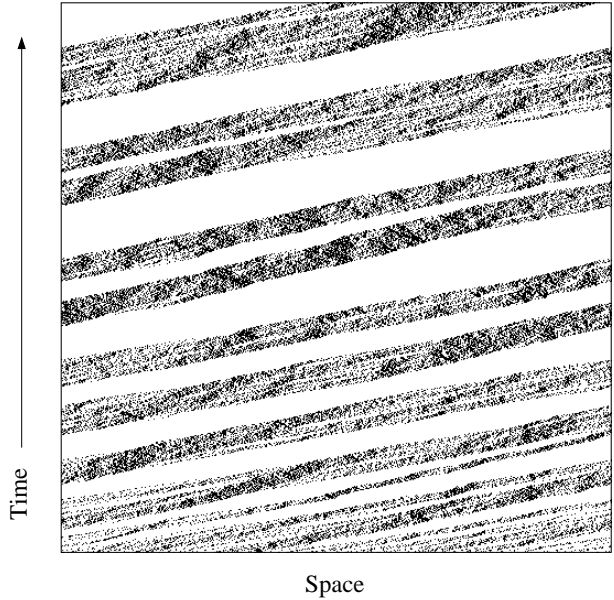}{80} 
	\caption{Space-time plot of bus positions for $\lambda = 0.02$, $\rho =
	0.2$, $\beta = 0.5$ and $L=500$. There are 10 time-steps between each
	snapshot on the time axis. Initially, the buses are positioned randomly and
	there are no passengers.}
	\label{fig:ST-Jammed}
\end{figure}

From our study of the BRM we provide strong evidence, both numerical and
analytical, that a true jamming phase transition does occur, but only in the
limit $\lambda \to 0^+$ with $\lambda L \to \infty$. The transition is from a
low density ``jammed'' phase to a high density homogeneous phase. When
$\lambda$ is small but finite, we find two strong signatures of the
transition. Firstly, the transition is rounded to a crossover; but this is
exponentially sharp in $1/\lambda$. Secondly, apparent coarsening occurs where
over long time scales, the system separates into jammed regions of finite size
with long but finite lifetimes. 

\begin{figure}
	\centps{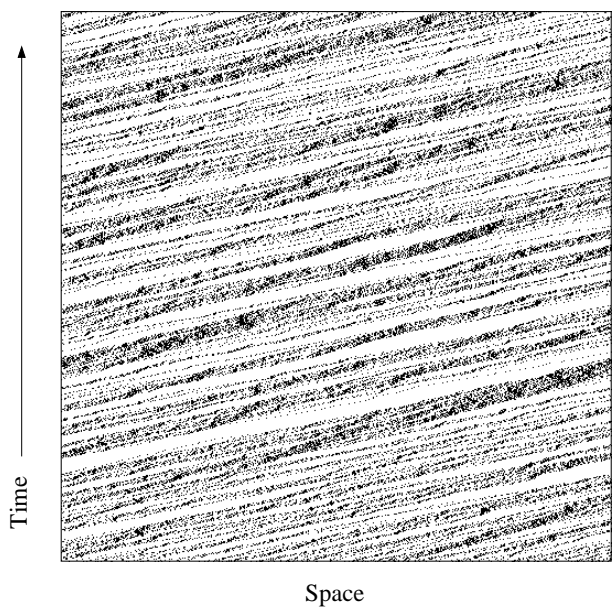}{80} 
	\caption{Space-time plot of bus positions for the same parameters as in
	fig.\ \ref{fig:ST-Jammed} with the exception that here, $\lambda = 0.1$.}
	\label{fig:ST-NoJam}
\end{figure}

\begin{figure}
	\centps{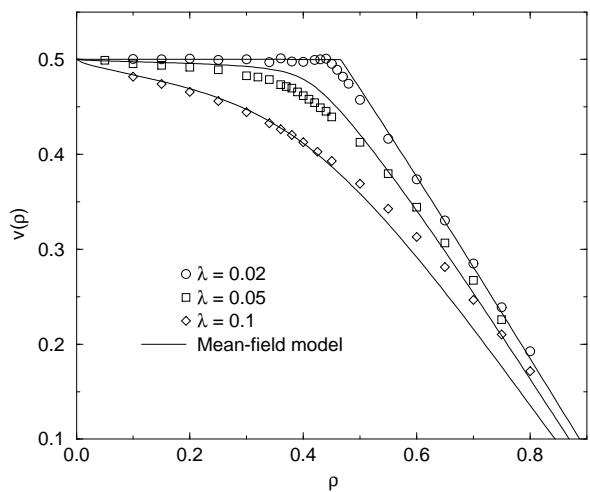}{90}
	\caption{The velocity as a function of bus density for $\beta = 0.5$ and
	various values of $\lambda$. The symbols are simulation results for the BRM
	with $L=10000$ and the lines are mean-field model (see below) results in
	the thermodynamic limit.}
	\label{fig:V-Rho}
\end{figure}

We first present some simulation results for the BRM. Figure
\ref{fig:ST-Jammed} shows a space-time plot of the system at low density and
small $\lambda$. As passengers enter the system, one sees the large inter-bus
gaps increasing in size until the system comprises several distinct clusters
(or ``jams'') of buses. The system then coarsens via coalescence of the bus
clusters until finally, only a single large cluster remains. For high
densities, we find that the system is homogeneous -- a snapshot of the system
as whole resembles the high density final cluster in
fig.\ \ref{fig:ST-Jammed}. Figure \ref{fig:ST-NoJam} shows a space-time plot for
the same system as in fig.\ \ref{fig:ST-Jammed}, with the exception that now
$\lambda = 0.1$. While small, transient clusters of buses do appear, the
``phase-separation'' seen for $\lambda = 0.02$ does not occur. Figure
\ref{fig:V-Rho} shows plots of bus velocity $v$ (average rate of hopping
forward) against bus density $\rho$. For the two larger values of $\lambda$,
velocity decreases smoothly with increasing density. However, for $\lambda =
0.02$, $v(\rho)$ has an {\it apparent} cusp at an intermediate value of the
density, suggesting the presence of a phase transition.

We now show that the BRM exhibits a phase transition in the limit $\lambda \to
0$ with $\lambda L \to \infty$. To see this, consider a system comprising a
single large cluster (as in fig.\ \ref{fig:ST-Jammed}). If $\lambda L \to
\infty$, then the site in front of the leading bus has passengers with
probability one (because the time since that site was last visited by a bus is
$\propto L$). Hence, the leading bus hops forward with probability
$\beta$. Since all of the gaps {\em within} the cluster are finite, there are
no passengers within the cluster as $\lambda \to 0$; the buses within the
cluster hop with probability one into unoccupied sites. The velocity (average
rate of hopping forward) of these buses is $1 - \rho_c$\cite{OEC}, where
$\rho_c$ is the density of buses in the cluster. For the cluster to be stable,
this velocity must equal that of the leading bus and so we have $\rho_c = 1 -
\beta$. For overall bus densities greater than $\rho_c$, the system becomes
homogeneous with all gaps finite. Therefore, we identify $\rho_c$ as the
critical density.

This shows that the BRM exhibits a phase transition in the limit of $\lambda
\to 0$. We now present a two-particle approximation to the problem which
suggests that there is no strict transition for non-zero $\lambda$. First, let
us approximate the probability that a bus hops into a gap of size $x$ by $u(x)
= f(x) + \beta(1-f(x))$, where $f(x)$ is an estimate of the probability that
there are no passengers on the first site of the gap. The average time since a
bus last left this site is $x/v$ (where $v$ is the average velocity in the
system), so we estimate $f(x) = \exp(-\lambda x/v)$ to give
\begin{equation}
u(x) = \beta + (1-\beta)\exp(-\lambda x/v)\;\;\;
	\mbox{for}\;\;\;x >0
\label{eqn:HopProb}
\end{equation}
with exclusion requiring $u(0) = 0$. This is in the spirit of a mean-field
approximation for the BRM, the nature of which is to replace the ``induced''
interaction between buses (which is subject to stochastic variation) with a
deterministic one. 

Now consider a ``jammed'' system as in fig.\ \ref{fig:ST-Jammed}, with the
large gap in front of the leading bus in the cluster having size $kL$ (where
$k$ is independent of $L$). We denote the size of the gap between the leading
two buses by $x$ so that, using the mean-field hopping rate in
(\ref{eqn:HopProb}), we may write a Langevin equation for the dynamics of this
gap size as
\begin{equation}
	\dot{x} = u(kL) - u(x) + \eta(t) 
	\equiv -\frac{d \Phi}{dx} + \eta(t).
	\label{eqn:Langevin}
\end{equation}
where $\eta(t)$ is a noise term (say white noise of unit
variance\cite{NoiseVariance}). The gap size $x$ has the dynamics of a particle
diffusing in a potential well $\Phi(x)$ given by (\ref{eqn:HopProb},
\ref{eqn:Langevin}). The potential has a maximum at $x^* = kL$ so that when $x
> x^*$, the particle has escaped from the well, or equivalently, the leading
bus has left the cluster. We denote the average time for this break-up to occur
by $\tau$, which is given by $\exp[\Phi(x^*) -
\Phi(0)]$ to a good approximation\cite{Kramers}. In the limit $L \to \infty$,
this becomes
\begin{equation}
	\tau \sim \exp \left[ \frac{\beta (1-\beta)}{\lambda}
	\right]
\label{tau2}
\end{equation}
which is finite for $\lambda > 0$, implying that a jam is not a stable object
and will eventually break up. However, when $\lambda \to 0$, the jam becomes
stable in agreement with our previous argument. When $\lambda$ is small but
non-zero, $\tau$ is exponentially large in $1/\lambda$ and it can appear that a
jam is stable when in fact it has a finite lifetime. Thus, we do not expect
true phase-separation to occur for non-zero $\lambda$.

Let us now move beyond the two-particle picture described above. Consider a
model of hopping particles where the hopping rate of a particle is a function
$u(x)$ of the size of the gap $x$ in front of that particle. By using the
mean-field expression for $u(x)$ given in (\ref{eqn:HopProb}), one defines a
new model which we call the mean-field model (MFM). The (rigorous) solution and
analysis of the steady state\cite{ZeroRange} of the MFM can be found in
\cite{OEC}; here we present some selected results.

The MFM exhibits no phase transition for non-zero $\lambda$ in agreement with
our two-particle argument but there is indeed a transition in the limit
$\lambda \to 0$ with $\lambda L \to \infty$. Figure \ref{fig:V-Rho} compares
velocity as a function of density in the MFM (solved analytically) and the BRM
(simulated); the agreement is quite good.  For $\lambda = 0.02$ in both models,
$v(\rho)$ has an {\em apparent} cusp at an intermediate value of the
density. We know that for the MFM, $v(\rho)$ is in fact non-singular since
there is rigorously no transition for non-zero $\lambda$. Since we believe that
the MFM captures the essential physics of the BRM, we expect that likewise
there is no transition for non-zero $\lambda$ in the BRM. When $\lambda$ is
small there is, however, a very sharp crossover between a low density
``jammed'' regime with $v \simeq \beta$, and a high density ``congested''
regime where $v$ decreases roughly linearly with increasing density.

To quantify the sharpness of the crossover for $\lambda$ close to zero in the
MFM, we calculated $\kappa_{max}$, the maximum curvature of $v(\rho)$.  For
$\lambda$ small (less than about $0.02$), we found\cite{OEC} that
$\kappa_{max}$ varies as $\exp(a/\lambda)$, where $a$ depends on
$\beta$. Therefore, although a strict phase transition occurs only in the limit
$\lambda \to 0$, the crossover is exponentially sharp in $1/\lambda$ for
small $\lambda$.

We now comment on the occurrence of apparent coarsening (see fig.\
\ref{fig:ST-Jammed}) in a system which, according to the above discussion, does
not strictly phase-separate. (On the one hand, we have argued that large
clusters are ultimately unstable while on the other hand, fig.\
\ref{fig:ST-Jammed} appears to show a fully phase-separated system.) We
believe\cite{OEC} that sufficiently large systems coarsen up to some finite
length scale which is exponentially large in $1/\lambda$. For the system in
fig.\ \ref{fig:ST-Jammed}, this length scale is much larger than the system
size.

\begin{figure}
	\centps{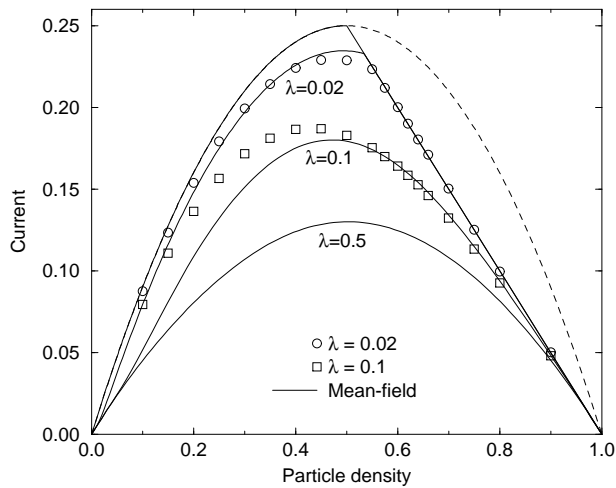}{90}
	\caption{Current as a function of density for the dual model for several
	values of $\lambda$. $L = 10000$ for all simulation data and MFM results
	are in the thermodynamic limit. The uppermost solid curve is the MFM result
	in the limit $\lambda \to 0$ with $\lambda L \to \infty$. The dashed curve
	is the exact result when $\lambda$ is set equal to zero before the
	thermodynamic limit is taken. The latter two curves are identical for $\rho
	< 0.5$.}
	\label{fig:Fundamental}
\end{figure}

Let us now return to the dual model defined earlier and interpret our findings
in that context. Since an inter-bus gap in the BRM corresponds to a cluster of
particles in the dual model, jamming is now a {\em high density} phenomenon,
characterised by the presence of large clusters of particles. This restores to
the word ``jamming'' a meaning closer to that used in everyday life. In the
limit $\lambda\to 0$, a phase transition arises from a low density homogeneous
phase in which the particles move quickly, to a high density jammed phase which
is characterised by macroscopic clusters of particles and a slow flow. An
infinitesimal rate $\lambda$ can result in macroscopic inhomogeneity and
decrease in flow. This is illustrated in fig.\ \ref{fig:Fundamental} which
shows the current (velocity times density) as a function of particle density
for the dual model.

A different interpretation of the dual model is as a model of stop-start
traffic flow with the particles representing cars. The longer a car is at rest,
the more likely it is that the driver will be slow to react when it is possible
to move again. This is related to several ``slow-to-start'' cellular automaton
traffic models studied recently\cite{SS}.

In conclusion, we have found that the BRM exhibits a jamming transition from a
high density homogeneous phase to a low density jammed phase. There is a
spontaneously broken symmetry in the jammed phase: one bus is selected over all
others to head the jam, even though all buses are identical. We have argued,
however, that a strict phase transition occurs only in the limit $\lambda \to
0$ with $\lambda L \to \infty$ and that for non-zero $\lambda$, one sees
crossover behaviour which is exponentially sharp in $1/\lambda$. Thus the model
exhibits an essential singularity at $\lambda = 0$ which causes, alongside the
dramatic crossover, the transient coarsening behaviour observed (see fig.\
\ref{fig:ST-Jammed} and \cite{OEC}) for small, positive $\lambda$. If similar 
phenomena were to arise in other models, this could easily be interpreted as
signifying a true phase transition where in fact none exists. Such phenomena
may indeed arise in certain cellular automata models of traffic\cite{Traffic}.

OJO is supported by a University of Edinburgh Postgraduate Research
Studentship. MRE is a Royal Society University Research Fellow.


\begin{thebibliography}{99}
\bibitem{MultiPassengers} While we have taken the passenger variable $\phi_i$
to be binary, this does not to forbid the presence of more than one passenger
at a site; we merely require that the extra passengers have no further effect
on the dynamics.
\bibitem{Overtaking} This is because the interchange of a fast-moving bus with
a slower-moving bus in front also interchanges their velocities. This contrasts
sharply with $1d$ models where jamming is induced by quenched
disorder\cite{defect}.
\bibitem{RealBuses} The relevance of the BRM to real buses is discussed further
in \cite{OEC}.
\bibitem{ExactMapping} In an exact mapping from the dual model to the the BRM,
a particle attempts to hop to the left when the site to its left is
updated. However, a more natural dynamics with no significant difference in
behaviour is to update only particles \cite{OEC}.
\bibitem{Galilean} Note that this violates Galilean invariance, requiring a 
pipe or some other fixed reference frame to be present.
\bibitem{NoiseVariance} The variance of the noise should strictly depend on
$\beta$ but since we are primarily interested in the effect of $\lambda$ on the
dynamics of the gap, we ignore this dependence.
\bibitem{ZeroRange} The MFM is an example of a zero range process for which
the steady state is given by a product measure.
\end{thebibliography}
\end{document}